\documentclass{article}
\usepackage{spconf,amsmath,graphicx}
\usepackage{color}
\usepackage{xcolor}
\usepackage{textgreek}
\usepackage{amssymb}
\usepackage{amsthm}
\usepackage{diagbox}
\usepackage{hyperref}
\usepackage{multirow}
\usepackage{siunitx}
\usepackage{balance}

\newcommand{\comment}[1]{}
\newcommand{\resolved}[1]{}

\newcommand{\D}{D}
\newcommand{\Loss}{\mathcal{L}}
\newcommand{\K}{K} 
\newcommand{\Exp}{E} 

\title{Real-time Speech Frequency Bandwidth Extension}
\name{Yunpeng Li \hspace{0.5cm} Marco Tagliasacchi \hspace{0.5cm} Oleg Rybakov \hspace{0.5cm} Victor Ungureanu \hspace {0.5cm} Dominik Roblek}
\address{
Google Research\\
{\tt\small\{yunpeng,mtagliasacchi,rybakov,ungureanu,droblek\}@google.com}
}
\begin{document}
%
\maketitle
\begin{abstract}
In this paper we propose a lightweight model for frequency bandwidth extension of speech signals, increasing the sampling frequency from 8kHz to 16kHz while restoring the high frequency content to a level almost indistinguishable from the 16kHz ground truth. The model architecture is based on SEANet (Sound EnhAncement Network), a wave-to-wave fully convolutional model, which uses a combination of feature losses and adversarial losses to reconstruct an enhanced version of the input speech. In addition, we propose a variant of SEANet that can be deployed on-device in streaming mode, achieving an architectural latency of 16ms. When profiled on a single core of a mobile CPU, processing one 16ms frame takes only 1.5ms. The low latency makes it viable for bi-directional voice communication systems.

\end{abstract}
%

\section{Introduction}
\label{sec:intro}

Frequency bandwidth reduction is one of the artifacts produced by lossy speech codecs used throughout telecommunication networks, leading to narrowband speech with frequencies up to 4kHz. This gives the received speech its familiar dull, muffled character. More recent coding standards, such as for example AMR-WB (Adaptive Multi-Rate Wideband), are explicitly designed to overcome this issue, operating in native wideband mode --- also known as HD Voice --- retaining frequencies up to 7kHz. However, to establish an HD Voice call, several requirements need to be met simultaneously: the mobile handsets, the underlying telephone infrastructure and the mobile cells all need to support wideband speech, and in some cases both the transmitter and the receiver need to be served by the same operator. As a consequence, HD voice calls are not yet ubiquitous, thus creating an opportunity for low-latency solutions deployed on the receiver device that aim at enhancing the quality of speech, reconstructing the high frequency content that was lost during transmission. 

In this paper we propose such a solution, devising a wave-to-wave model for frequency bandwidth extension, which is fully causal and sufficiently lightweight for real-time applications on mobile devices. Our model is trained using a combination of adversarial and reconstruction objectives, defined by a discriminator learned concurrently with the generator, with a design similar to the audio-only version of SEANet (Sound Enhancement Network) that we introduced in our previous work~\cite{tagliasacchi2020}.

In our experiments we mainly target the case of increasing the sampling frequency from 8kHz to 16kHz, thus extending the available bandwidth from 4kHz to 8kHz, respectively. We evaluate the output of the model with both objective evaluation metrics and subjective ratings, and we observe that the quality is indistinguishable from the original samples at 16kHz.
We further investigate the scenario where the input bandwidth at test time does not exactly match what was used during training, where catastrophic failure can occur even for relatively small mismatches. 
We show that the model can be made robust against such mismatches by using a range of bandwidths during learning.
Moreover, and somewhat surprisingly, the robustness induced by such a learning scheme persists even when the input bandwidth at test time is outside of the range seen during training.

In addition, one of our main contributions is to explicitly address the problem of deploying SEANet on a mobile device, aiming at low latency. This is inspired by previous work on streaming architectures for keyword spotting~\cite{rybakov2020streaming}, that we extended to support the operations needed to deploy a UNet generator. With this solution, we are able to process each 16ms audio frame in $\sim$1.5ms on the CPU of a mobile device, so that the total latency is $\sim$17.5 ms. 

The problem of generating an audio waveform at a higher sampling frequency is known in the literature either as audio superresolution or frequency bandwidth extension. 
Early methods were based on the source-filter model of speech production and exploit DNNs to estimate the upper frequency envelope~\cite{abel2016}. Inspired by the early success in image superresolution~\cite{dong2016}, end-to-end audio-based solutions were proposed, based on wave-to-wave UNet~\cite{kuleshov2017superres}, WaveNet~\cite{wang2018, gupta2019}, hybrid time/frequency-domain models~\cite{lim2018}. All these methods are trained by minimizing a reconstruction loss, typically the $\ell_2$ norm in the time domain. 
In the field of audio denoising, better qualitative results are obtained computing the reconstruction loss in some feature space~\cite{germain2018senet}, and using adversarial losses~\cite{Pascual2017}, hence a similar approach has also been adopted for audio superresolution~\cite{eskimez2019,kim2019bandwidth}. In particular,~\cite{kim2019bandwidth} leverages a mix of adversarial and reconstruction losses, computed both in the original time domain and in the feature space defined by the bottleneck of a pre-trained autoencoder. 

The design of SEANet is similar to~\cite{kim2019bandwidth}, but we adopt the losses proposed in~\cite{kumar2019}, in which the reconstruction loss is computed in the feature space of the discriminator, at different scales and at different layers. In addition, we explicitly address the problem of low-delay inference, which was not addressed  in previous work in this area.

\section{Method}
\label{sec:method}

\subsection{Model Architecture}
\label{sec:architecture}

The generator is a 1D convolutional U-Net~\cite{unet} that adopts the same overall architecture as the audio-only SEANet proposed in~\cite{tagliasacchi2020}. 
The model is a symmetric encoder-decoder network with skip-connections and residual units.
The encoder and the decoder each have four convolution blocks, which are sandwiched between two plain convolution layers. The encoder follows a down-sampling scheme of (2, 2, 8, 8) while the decoder up-samples in the reverse order. The number of channels is doubled whenever down-sampling and halved whenever up-sampling.
Each encoder block consists of three residual units each containing convolutions with dilation rates of 1, 3, and 9, respectively, followed by a down-sampling layer in the form of a strided convolution. The decoder block mirrors the encoder block, and consists of a transposed convolution for up-sampling followed by the same three residual units.
A skip-connection is added between each encoder block and its mirrored decoder block. The out-most skip-connection directly connects the input waveform to the output waveform.
We refer interested readers to~\cite{tagliasacchi2020} for details of the SEANet architecture.

To be able to process live audio streams in real-time and potentially on low-power mobile devices, our model differs from the original SEANet version in several ways:
\begin{itemize}
    \item All convolutions are \emph{causal}. This means that padding is only applied to the past but not the future in training and offline inference, whereas no padding is used in streaming inference.
    \item The model is much more lightweight. We reduce the number of channels at each layer 4-fold, which results in a model approximately 1/16 the size and computational cost of the original model.
    \item We do not apply any normalization. While weight normalization~\cite{salimans2016} is beneficial in the original model, it did not help in the smaller model likely due to having much fewer weights in the convolution kernels.
\end{itemize}
We call this new model the \emph{Streaming SEANet} to distinguish it from the original version, and we will describe in Section~\ref{sec:streaming} how real-time processing can be realized by means of streaming convolutions.

As in the original SEANet model, we use the same multi-resolution convolutional discriminator proposed in~\cite{kumar2019}.
Three structurally identical discriminators are applied to the input audio at different resolutions: original, 2-times down-sampled, and 4-times down-sampled.
Each discriminator consists of an initial plain convolution followed by four grouped convolutions~\cite{krizhevsky2012}, each of which has a group size of 4, a down-sampling factor of 4, and a channel multiplier of 4 up to a maximum of 1024 output channels. They are followed by two more plain convolution layers to produce the final output, i.e., the logits. Since the discriminator is fully convolutional, the number of logits in the output is proportional to the length of the input audio. 
See~\cite{kumar2019} for more architectural details of the discriminator.

We use ELU activation~\cite{clevert2016} in the generator and Leaky ReLU activation~\cite{maas2013} with the default $\alpha=0.2$ in the discriminator.
All residual and skip connections in the generator are pre-activation~\cite{he2016}.
Layer normalization~\cite{ba2016} is used in the discriminator whereas the generator does not use any normalization.

\subsection{Learning}
\label{sec:learning}

We use the same adversarial and reconstruction objective as~\cite{tagliasacchi2020}, albeit without the auxiliary accelerometer signal. 
Given a time-aligned input-target audio pair $(x, y)$, let $G(x)$ be the output audio of the generator.
The adversarial loss is a hinge loss over the logits of the discriminator, averaged over multiple resolutions and over time.
More formally, let $k\in\{1,\dots,K\}$ index over the individual discriminators for different resolutions ($K=3$ in our case), and $t$ index over the length of the output, i.e., the number of logits $T_k$, of the discriminator at scale $k$. The discriminator is trained to classify clean vs. generated audio, by minimizing
\begin{align}
    \nonumber
    \Loss_\D =~& \Exp_{y} \left [\frac{1}{\K}\sum_{k} \frac{1}{T_k}\sum_{t} \max(0, 1 - \D_{k,t}(y)) \right ] + \\  
    & \Exp_{x} \left [ \frac{1}{\K}\sum_{k} \frac{1}{T_k}\sum_{t} \max(0, 1 + \D_{k,t}(G(x)) \right ],  
\end{align}
while the adversarial loss for the generator is
\begin{equation}
    \Loss_G^{\text{adv}} = \Exp_{x} \left [ \frac{1}{\K}\sum_{k,t} \frac{1}{T_k} \max(0, 1 - \D_{k,t}(G(x)) \right ].  
\end{equation}
The reconstruction loss is the ``feature'' loss proposed in~\cite{kumar2019}, namely the average absolute difference between the discriminator's internal layer outputs for the generated audio and those for the corresponding target audio.
\begin{equation}
    \Loss_G^{\text{rec}} = \Exp_{x} \left [ \frac{1}{KL}\sum_{k,l} \frac{1}{T_{k,l}} \sum_{t} \left| \D_{k,t}^{(l)}(y) - \D_{k,t}^{(l)}(G(x)) \right| \right ],
\end{equation}
where $L$ is the number of internal layers, $D_{l,t}^{(l)}$ ($l\in\{1,\dots,L\}$) is the $t$-th output of layer $l$ of discriminator $k$, and $T_{k,l}$ denotes the length of the layer in the time dimension.
The overall generator loss is a weighted sum of the adversarial loss $\Loss_G^{\text{adv}}$ and the reconstruction loss $\Loss_G^{\text{rec}}$. 
%
%

In all our experiments, we train for 1 million steps with a batch size of 16 using the same optimizer parameters and a weighting factor of 100 for $\Loss_G^{\text{rec}}$ as in~\cite{tagliasacchi2020}.

\subsection{Real-time Processing of Streaming Input}  
\label{sec:streaming}


Unlike offline inference where the entire input audio is available at once, in streaming applications the input is fed continuously.
For efficient computation with low delay, convolutions must be performed in a streaming fashion where rolling buffers are used to maintain past intermediate outputs. 
A key challenge is to provide automated conversion between regular convolutions for training and streaming convolutions for inference under a common interface, as this allows the decoupling of high-level model architecture design from low-level considerations for streaming inference.

To this end, we extend the Streaming-aware Neural Network~\cite{rybakov2020streaming} to further support three types of operations: strided convolutions, transposed convolutions, and convolutions with shortcut connections. While the original framework already supports plain convolutions, this extension enables streaming convolutions for U-Net architectures. We omit the implementation details here in the interest of space, as the extended framework is open-source and freely available
at~\cite{OPEN}.

\textbf{Latency:}
Since all convolutions in our model are causal, the architectural latency results entirely from the presence of strided and transposed convolutions:
Given that the innermost layers have a temporal resolution $2\times2\times8\times8=256$-times coarser than the audio, the decoder can only produce output audio samples in chunks of 256. This translates to a latency of 16ms at 16KHz sampling rate. Profiling the model on a single CPU core of a Pixel 4 mobile phone indicates a processing time of 1.5ms for each 16ms chunk of audio, giving a total latency of 17.5ms.

\section{Experiments}
\label{sec:experiments}

We focus on extension of frequency bandwidth of sub-4KHz up to 8KHz (at a sampling rate of 16KHz), since the quality gain of speech audio beyond 8KHz is marginal in comparison with the 4KHz--8KHz range.
We evaluate our model on the widely-used VCTK dataset~\cite{vctk}, using the default training/testing split.
All the audio is resampled to 16KHz.


Two objective metrics are used to measure the reconstruction quality of the reconstructed audio with respect to the ground truth:
\begin{itemize}
    \item Scale-invariant signal-to-distortion ratio (SI-SDR)~\cite{si-sdr} over the audio waveform, which measures the per-sample fidelity up to a uniform scaling factor.
    \item VGG distance, the L2-distance between the ground truth and the reconstructed audio embeddings computed by a pre-trained \mbox{VGGish} network~\cite{hershey2017vggish}.
\end{itemize}
Since the VGGish network operates on mel-spectrograms and is trained on a wide range of audio classification tasks, the VGG distance is expected to be less sensitive to per-sample alignment than SI-SDR and more reflective of perceived audio quality.
For each model configuration, training is repeated 5 times so as to compute the mean and its standard error.

\subsection{Effects of Training with Variable-band Inputs}

\newcommand{\mas}[2]{{#1}{\footnotesize$\pm${#2}}}

\begin{table}
    \centering
    \begin{tabular}{|l|*{2}{c|}}\hline
        \backslashbox{Test band}{Train band}
        &\makebox[5em]{Medium}&\makebox[5em]{Variable}\\
        \hline
        \multirow{2}{*}{Wide (16.9dB, 1.73)}
            & \mas{-3.4}{0.4}dB & \mas{18.0}{0.1}dB \\
            & \mas{3.04}{0.03} & \mas{1.09}{0.01} \\
        \hline
        \multirow{2}{*}{Medium (9.2dB, 2.17)} 
            & \mas{17.4}{0.1}dB & \mas{15.3}{0.1}dB \\
            & \mas{1.13}{0.01} & \mas{1.18}{0.01} \\
        \hline
        \multirow{2}{*}{Narrow (3.8dB, 2.63)} 
            & \mas{4.7}{0.2}dB & \mas{10.9}{0.1}dB \\
            & \mas{1.91}{0.03} & \mas{1.39}{0.02} \\
        \hline
        \end{tabular}
    \caption{
        Performance of Streaming SEANet with different training and testing input frequency bands. Above: SI-SDR, higher is better. Below: VGG distance, lower is better. All values are the absolute (as opposed to relative). The average SI-SDR and VGG distance for the input itself is shown in the parentheses in the left-most column.
    }
    \label{tab:vckt-fixed-vs-var}
\end{table}


To assess the robustness of frequency extension, we evaluate on three slightly different input frequency bands: 100--3800Hz (``wide''), 200--3600Hz (``medium''), and 300--3400Hz (``narrow'').
Two types of models are trained based on frequency bands of their inputs during training: fixed ``medium'' band (200--3600Hz) and ``variable'' band, where the low and high frequency cutoffs are sampled uniformly in the ranges of [0, 300Hz] and [3400Hz, 4000Hz] respectively.
The input is produced from the ground truth audio by means of band-pass filtering.
For the variable-band model, frequency band sampling and band-pass filtering are performed on-the-fly during training akin to typical data augmentation approaches.

Table~\ref{tab:vckt-fixed-vs-var} shows the average SI-SDR and VGG distance respectively, of applying each model variant on each type of test input. We can see that while the model trained on fixed-band inputs performs slightly better on test inputs of the exact same type, it is highly susceptible to frequency band mismatch. In contrast, the variable-band model is able to maintain good performance over a range of test inputs. This shows that training with variable-band inputs makes the model much more robust.

Surprisingly, the robustness generalizes even to inputs with unseen frequency ranges. To illustrate this, we force the models to ``enhance'' the ground truth itself. As expected, the fixed-band model fails completely (with SI-SDR below -9dB and VGG distance over 3.3). The variable-band model, however, is able to avoid drastic degradation (maintaining SI-SDR around 15dB and VGG distance around 1.3) despite having never encountered this kind of input during training.

For the rest of the paper, we will assume models to be trained on variable-band inputs and tested on ``medium'' unless noted otherwise.

\subsection{Effects of Discriminator-based Learning Objectives}


\begin{figure}
    \newcommand{\spectrogramwidth}{0.12\textwidth}
    \newcommand{\boxwidth}{0.095\textwidth}
    \centering
    \begin{tabular}{cccc}
        \makebox[\boxwidth]{\includegraphics[width=\spectrogramwidth]{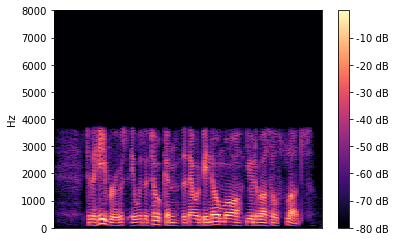}} &
        \makebox[\boxwidth]{\includegraphics[width=\spectrogramwidth]{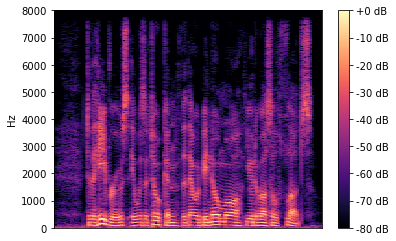}} &
        \makebox[\boxwidth]{\includegraphics[width=\spectrogramwidth]{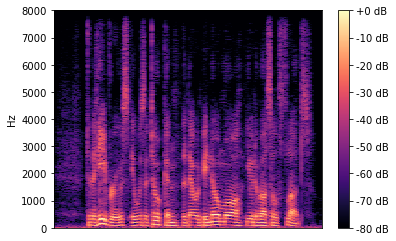}} &
        \makebox[\boxwidth]{\includegraphics[width=\spectrogramwidth]{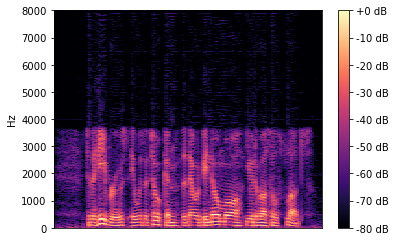}} \\
        (a) & (b) & (c) & (d) 
    \end{tabular}
    \caption{Example STFT spectrograms of: (a) bandwidth-compressed input (13.1dB, 1.78), (b) ground truth, (c) output of the adversarially-trained model (16.7dB, 1.16), (d) output of the discriminator-free variant (18.2dB, 2.22).}
    \label{fig:spectrograms}
\end{figure}

To understand the effects of having adversarial and reconstruction losses defined by the discriminator, we trained an alternative model where we use no discriminator, but only an $\ell_1$ reconstruction loss on the waveform. 
This model obtained an average SI-SDR of 18.1dB and VGG distance of 2.06, compared with 15.3dB and 1.18, respectively, by the adversarially-trained model.
Clearly, the discriminator-free variant achieves better SI-SDR at the cost of worse VGG distance.
The difference is further illustrated in Figure~\ref{fig:spectrograms}, where the STFT spectrograms indicate that the discriminator-free variant produces hardly any extension in the high-frequency range. This is confirmed by our aural inspection of its output samples, which are perceptually indistinguishable from the input. Therefore, we can make two observations: i) The discriminator and the losses it defines are important for frequency extension, and ii) SI-SDR can be very misaligned with perceived quality and much more so than VGG distance.

\subsection{Comparison with Existing Baselines}
\label{sec:baselines}

\begin{table}
    \centering
    \begin{tabular}{|l|*{2}{c|}}
        \hline
        \backslashbox[10em]{Model}{Metric}
        &\makebox[6em]{SI-SDR}&\makebox[6em]{VGG distance}\\
        \hline
        Streaming SEANet                             & \mas{15.3}{0.1}dB & \mas{1.18}{0.01} \\
        Offline SEANet~\cite{tagliasacchi2020}       & \mas{18.9}{0.1}dB & \mas{1.03}{0.03} \\
        Germain {\it et al.}~\cite{germain2018senet} & \mas{14.3}{0.9}dB & \mas{1.66}{0.04} \\
        \hline
        \end{tabular}
    \caption{Performance of Streaming SEANet and the offline baselines.}
    \label{tab:compare-vs-baselines}
\end{table}

We further compare our Streaming SEANet with two baselines, the original offline SEANet and the model by Germain {\it et al.}~\cite{germain2018senet}, both of which are more than an order of magnitude more expensive in computational cost, with architectural latency in the hundreds of milliseconds.
All models are trained with the same loss functions.
Table~\ref{tab:compare-vs-baselines} compares the performance of these models. The results show the streaming model achieves comparable performance to these baselines, despite being much more lightweight and low-latency.



\begin{figure}
    \centering
    \includegraphics[width=0.48\textwidth]{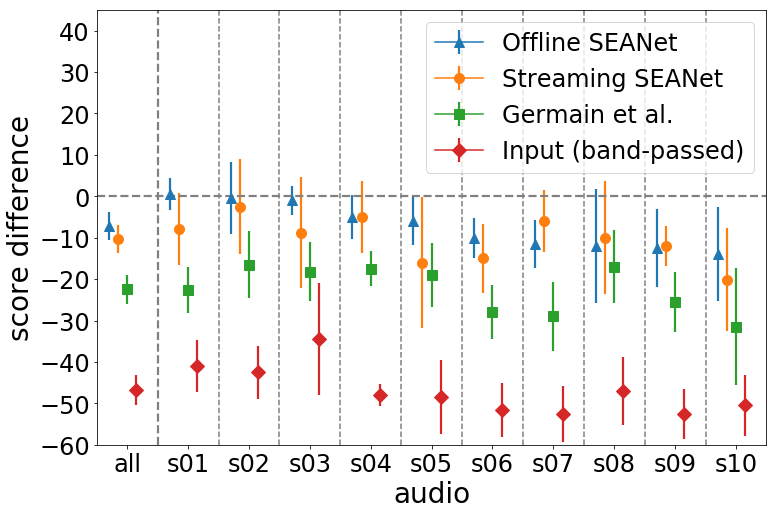}
    \caption{Average subjective score differences on VCTK, \emph{relative to clean audio}, together with 95\% confidence intervals. First column is the mean of each model over all audio clips.}
    \label{fig:subjective-eval}
\end{figure}

To assess perceptual quality, we perform subjective evaluation
using the MUSHRA methodology~\cite{mushra2015} with 10 VCTK test audio clips (2--5 seconds each). We compare the ground truth and the input (200--3600Hz band-passed), together with Streaming SEANet and the two baselines described above. This results in 10 groups of 5 audio clips, where clips within a group have the same speech content, but vary in quality. To calibrate our 10 raters, we first present them two band-passed examples (score~=~50) with their corresponding ground truth (score~=~100). We then ask the raters to assign scores between 0 and 100 to each of the $10 \times 5$ clips\footnote{\url{https://google-research.github.io/seanet/freqext/examples/vctk.html}}. We summarize the results in Figure~\ref{fig:subjective-eval} and confirm the quality ranking \textit{Ground truth $>$ Offline SEANet $>$ Streaming SEANet $>$ Germain et al. $>$ Input (band-passed)} is statistically significant, using the Wilcoxon signed-rank test 
(p-value $=0.0038$ between Offline and Streaming SEANet, and $<10^{-6}$ everywhere else). 
We note this ranking is consistent with the quantitative metrics in Table~\ref{tab:compare-vs-baselines}, and that there exists a relatively small gap 
between Streaming SEANet and the offline SEANet.

\section{Conclusion}
\label{sec:conclusion}

We presented a wave-to-wave model for frequency bandwidth extension, and showed that it can be made robust to unexpected out-of-distribution inputs by varying input bandwidth during training.
Our model is lightweight and fully causal, and yet attains quality levels comparable to those of much larger offline models. 
With our extended support for streaming convolution, the model is capable of processing live audio in real time on ordinary mobile devices.

\balance
\bibliographystyle{IEEEtran}
{
\small  
\bibliography{references}

\begin{thebibliography}{10}
\providecommand{\url}[1]{#1}
\csname url@samestyle\endcsname
\providecommand{\newblock}{\relax}
\providecommand{\bibinfo}[2]{#2}
\providecommand{\BIBentrySTDinterwordspacing}{\spaceskip=0pt\relax}
\providecommand{\BIBentryALTinterwordstretchfactor}{4}
\providecommand{\BIBentryALTinterwordspacing}{\spaceskip=\fontdimen2\font plus
\BIBentryALTinterwordstretchfactor\fontdimen3\font minus
  \fontdimen4\font\relax}
\providecommand{\BIBforeignlanguage}[2]{{%
\expandafter\ifx\csname l@#1\endcsname\relax
\typeout{** WARNING: IEEEtran.bst: No hyphenation pattern has been}%
\typeout{** loaded for the language `#1'. Using the pattern for}%
\typeout{** the default language instead.}%
\else
\language=\csname l@#1\endcsname
\fi
#2}}
\providecommand{\BIBdecl}{\relax}
\BIBdecl

\bibitem{tagliasacchi2020}
M.~Tagliasacchi, Y.~Li, K.~Misiunas, and D.~Roblek, ``{SEANet}: A multi-modal
  speech enhancement network,'' in \emph{INTERSPEECH}, 2020.

\bibitem{rybakov2020streaming}
\BIBentryALTinterwordspacing
O.~Rybakov, N.~Kononenko, N.~Subrahmanya, M.~Visontai, and S.~Laurenzo,
  ``Streaming keyword spotting on mobile devices,'' \emph{CoRR}, 2020.
  [Online]. Available: \url{https://arxiv.org/abs/2005.06720}
\BIBentrySTDinterwordspacing

\bibitem{abel2016}
J.~{Abel}, M.~{Strake}, and T.~{Fingscheidt}, ``Artificial bandwidth extension
  using deep neural networks for spectral envelope estimation,'' in \emph{2016
  IEEE International Workshop on Acoustic Signal Enhancement (IWAENC)}, 2016,
  pp. 1--5.

\bibitem{dong2016}
C.~{Dong}, C.~C. {Loy}, K.~{He}, and X.~{Tang}, ``Image super-resolution using
  deep convolutional networks,'' \emph{IEEE Transactions on Pattern Analysis
  and Machine Intelligence}, vol.~38, no.~2, pp. 295--307, 2016.

\bibitem{kuleshov2017superres}
\BIBentryALTinterwordspacing
V.~Kuleshov, S.~Z. Enam, and S.~Ermon, ``Audio super resolution using neural
  networks,'' \emph{CoRR}, vol. abs/1708.00853, 2017. [Online]. Available:
  \url{http://arxiv.org/abs/1708.00853}
\BIBentrySTDinterwordspacing

\bibitem{wang2018}
M.~{Wang}, Z.~{Wu}, S.~{Kang}, X.~{Wu}, J.~{Jia}, D.~{Su}, D.~{Yu}, and
  H.~{Meng}, ``Speech super-resolution using parallel wavenet,'' in \emph{2018
  11th International Symposium on Chinese Spoken Language Processing (ISCSLP)},
  2018, pp. 260--264.

\bibitem{gupta2019}
A.~{Gupta}, B.~{Shillingford}, Y.~{Assael}, and T.~C. {Walters}, ``Speech
  bandwidth extension with wavenet,'' in \emph{2019 IEEE Workshop on
  Applications of Signal Processing to Audio and Acoustics (WASPAA)}, 2019, pp.
  205--208.

\bibitem{lim2018}
T.~Y. {Lim}, R.~A. {Yeh}, Y.~{Xu}, M.~N. {Do}, and M.~{Hasegawa-Johnson},
  ``Time-frequency networks for audio super-resolution,'' in \emph{2018 IEEE
  International Conference on Acoustics, Speech and Signal Processing
  (ICASSP)}, 2018, pp. 646--650.

\bibitem{germain2018senet}
F.~G. Germain, Q.~Chen, and V.~Koltun, ``Speech denoising with deep feature
  losses,'' 2018.

\bibitem{Pascual2017}
S.~Pascual, A.~Bonafonte, and J.~Serrà, ``{SEGAN}: Speech enhancement
  generative adversarial network,'' in \emph{INTERSPEECH}, 2017, pp.
  3642--3646.

\bibitem{eskimez2019}
S.~E. {Eskimez}, K.~{Koishida}, and Z.~{Duan}, ``Adversarial training for
  speech super-resolution,'' \emph{IEEE Journal of Selected Topics in Signal
  Processing}, vol.~13, no.~2, pp. 347--358, 2019.

\bibitem{kim2019bandwidth}
S.~Kim and V.~Sathe, ``Bandwidth extension on raw audio via generative
  adversarial networks,'' 2019.

\bibitem{kumar2019}
K.~Kumar, R.~Kumar, T.~de~Boissiere, L.~Gestin, W.~Z. Teoh, J.~Sotelo,
  A.~de~Brebisson, Y.~Bengio, and A.~Courville, ``{MelGAN}: Generative
  adversarial networks for conditional waveform synthesis,'' in \emph{Advances
  in Neural Information Processing Systems}, 2019.

\bibitem{unet}
O.~Ronneberger, P.~Fischer, and T.~Brox, ``U-net: Convolutional networks for
  biomedical image segmentation,'' in \emph{Medical Image Computing and
  Computer-Assisted Intervention -- MICCAI 2015}, N.~Navab, J.~Hornegger, W.~M.
  Wells, and A.~F. Frangi, Eds.\hskip 1em plus 0.5em minus 0.4em\relax Cham:
  Springer International Publishing, 2015, pp. 234--241.

\bibitem{salimans2016}
T.~Salimans and D.~P. Kingma, ``Weight normalization: A simple
  reparameterization to accelerate training of deep neural networks,'' in
  \emph{Advances in Neural Information Processing Systems}, 2016, pp. 901--909.

\bibitem{krizhevsky2012}
A.~Krizhevsky, I.~Sutskever, and G.~E. Hinton, ``Imagenet classification with
  deep convolutional neural networks,'' in \emph{Advances in Neural Information
  Processing Systems}, 2012, pp. 1097--1105.

\bibitem{clevert2016}
D.-A. Clevert, T.~Unterthiner, and S.~Hochreiter, ``Fast and accurate deep
  network learning by exponential linear units ({ELUs}),'' in
  \emph{International Conference on Learning Representations}, 2016.

\bibitem{maas2013}
A.~L. Maas, A.~Y. Hannun, and A.~Y. Ng, ``Rectifier nonlinearities improve
  neural network acoustic models,'' in \emph{ICML Workshop on Deep Learning for
  Audio, Speech and Language Processing}, 2013.

\bibitem{he2016}
\BIBentryALTinterwordspacing
K.~He, X.~Zhang, S.~Ren, and J.~Sun, ``Identity mappings in deep residual
  networks,'' \emph{CoRR}, vol. abs/1603.05027, 2016. [Online]. Available:
  \url{http://arxiv.org/abs/1603.05027}
\BIBentrySTDinterwordspacing

\bibitem{ba2016}
J.~L. Ba, J.~R. Kiros, and G.~E. Hinton, ``Layer normalization,'' \emph{arXiv
  preprint arXiv:1607.06450}, 2016.

\bibitem{OPEN}
``Streaming aware neural network models,''
  \url{https://github.com/google-research/google-research/tree/master/kws_streaming}.

\bibitem{vctk}
C.~Veaux, J.~Yamagishi, and K.~MacDonald, ``{CSTR VCTK Corpus}: English
  multi-speaker corpus for cstr voice cloning toolkit,'' 2016.

\bibitem{si-sdr}
J.~L. {Roux}, S.~{Wisdom}, H.~{Erdogan}, and J.~R. {Hershey}, ``{SDR} –
  half-baked or well done?'' in \emph{ICASSP 2019 - 2019 IEEE International
  Conference on Acoustics, Speech and Signal Processing (ICASSP)}, 2019, pp.
  626--630.

\bibitem{hershey2017vggish}
\BIBentryALTinterwordspacing
S.~Hershey, S.~Chaudhuri, D.~P.~W. Ellis, J.~F. Gemmeke, A.~Jansen, C.~Moore,
  M.~Plakal, D.~Platt, R.~A. Saurous, B.~Seybold, M.~Slaney, R.~Weiss, and
  K.~Wilson, ``{CNN} architectures for large-scale audio classification,'' in
  \emph{International Conference on Acoustics, Speech and Signal Processing
  (ICASSP)}, 2017. [Online]. Available: \url{https://arxiv.org/abs/1609.09430}
\BIBentrySTDinterwordspacing

\bibitem{mushra2015}
``Method for the subjective assessment of intermediate quality levels of coding
  systems,'' ITU-Recommendation 655 BS.1534-3, 2015.

\end{thebibliography}
}

\end{document}